# NUMEXO2: A versatile digitizer for Nuclear Physics

**C.Houarner[1,a], A.Boujrad[a], M.Tripon[a], M.Bezard[a], M.Blaizot[a], P.Bourgault[a], S.Coudert[a], B.Duclos[a], F.J.Egea[b], G de France[a], A.Gadea[b], A.Lemasson[a], L.Martina[a], C.Maugeais[a], J.Pancin[a], B.Raine[a], F.Saillant[a], A.Triossi[c] and G.Wittwer[a]**

[a] Grand Accélérateur National d'Ions Lourds (GANIL), CEA/DRF-CNRS/IN2P3, Bvd Henri Becquerel, 14076 Caen, France
[b] Instituto de Física Corpuscular, CSIC-Universidad de Valencia, Edificio Institutos de Investigacion c/ Catedratico Jose Beltran 2, E-46980 Paterna (Valencia), Spain
[c] Università degli Studi di Padova and INFN, Sezione di Padova, Via F. Marzolo 8, 35131, Padova, Italy

E-mail: charles.houarner@ganil.fr

ABSTRACT: NUMEXO2 is a 16 channels 14bit/200MHz digitizer and processing board initially developed for gamma-ray spectroscopy (for EXOGAM: EXOtic nuclei GAMma ray). Numexo2 has been gradually extended and improved as a general purpose digitizer to fulfill various needs in nuclear physics detection at GANIL. This was possible thanks to reprogrammable components like FPGAs and the optimization of different algorithms. The originality of this work compared to similar systems is that all numerical operations follow the digital data flow from ADCs, without any storage step of samples. Some details are given on digital processing of the signals, delivered by a large variety of detectors: HPGe, silicon strip detector, ionisation chamber, liquid and plastic scintillators read-out with photomultipliers, Multi Wire Proportional Counter and drift chamber. Thanks to this high versatility, the NUMEXO2 digitizer is extensively used at GANIL (Grand Accélérateur National d'Ions Lourds). Some of the performances of the module are also reported.

---

[1] Corresponding author.

# Content

# 1. Introduction

Data acquisition systems for experimental nuclear physics are designed to collect, digitize, process, format and transmit the data produced by the particle detectors. In complex systems, with several detectors, usually also includes a time synchronization system in order to build the full events from the sub-events delivered by each individual detector. The requirements for such a data acquisition system are (i) small dead-time even at high counting rates, as well as having (ii) good energy and time resolutions, in a wide energy range with low energy threshold. The use of new generation of ADC allows size and power reduction and enables new possibilities in nuclear physics experiments thanks to the flexibility offered by digital approach including various PSA (Pulse Shape Analysis) capabilities. Moreover, the parallel processing capabilities of FPGAs enables a higher throughput and further reduces the event processing time, thanks to several Digital Signal Processing units inside the FPGA and large amount of logic cells. The NUMEXO2 (NUMériseur pour EXOgam2) digitizer was initially developed for the EXOGAM (EXOtic nuclei GAMma ray) array [1] by a large international collaboration and its extension to other applications was made internally at GANIL (Grand Accélérateur National d'Ions Lourds).

# 2. The NUMEXO2 board

## 2.1 General description

The NUMEXO2 is a 16 channels electronic module on NIM (Nuclear Instrumentation Module) standard [2]. It was initially designed for EXOGAM detector to replace the obsolete VXI digitizers based on analog architecture. It was gradually extended and improved as a general purpose digitizer to fulfill various needs in nuclear physics detection at GANIL. The NUMEXO2 boards have been adapted to a large variety of detectors such as Double Sided Silicon Detectors, $BaF_2$, Multi-Wire Proportional Chamber, Ionization Chamber and PhotoMultiplier... More than 120 NUMEXO2 modules were manufactured for GANIL needs. Today, NUMEXO2 is a multifunction module working at high counting rate (up to 50 kHz per Channel), able to perform Digital Signal Processing, Time measurements and Pulse Shape Analysis. In addition, some external modules as TAC (Time to Amplitude Converter) or MPD4 (Multi Pulse shape Discriminator) outputs can also be processed in NUMEXO2.

## 2.2 Mother and daughters boards design

NUMEXO2 module is based on one mother board and four daughter boards. The motherboard is a 16-layer class 7 FR-4 printed circuit board. It holds two large FPGAs: VIRTEX6 family (LX130T) and VIRTEX5 family (FX70T) from Xilinx manufacturer. The digitized data are transmitted to the first FPGA (VIRTEX6) to be filtered and processed by algorithms calculating energy and time information of the detected signals. The calculated parameters are embedded in a reference frame by a PPC440 processor running Linux, and then processed by a second FPGA (VIRTEX5) in charge of associating a universal Time Stamp (TS) and outputting the data through Gigabit Ethernet or optical fiber to acquisition or analysis computers. This Time Stamp is generated by the Global Trigger and Synchronization (GTS) system [3], that also distributes a common 100 MHz clock and broadcasts the Time Stamp to each board in the system. Some details on the GTS are given in 3.2.2.

The four daughter boards, each with 4 digitizer channels, are dedicated to Analog to Digital Conversion operations. Each board includes two dual 14 bits/250MHz Texas Instruments ADC (ADS62P49) downscaled to 200 MHz, as a multiple of the GTS frequency. The daughter card holds a dedicated PLL to minimize the clock jitter, which is a critical parameter on ADC resolution [4]. Each channel is independent and has its own dead-time as it is a triggerless system. Figure 1 shows the block-diagram of the NUMEXO2 board.

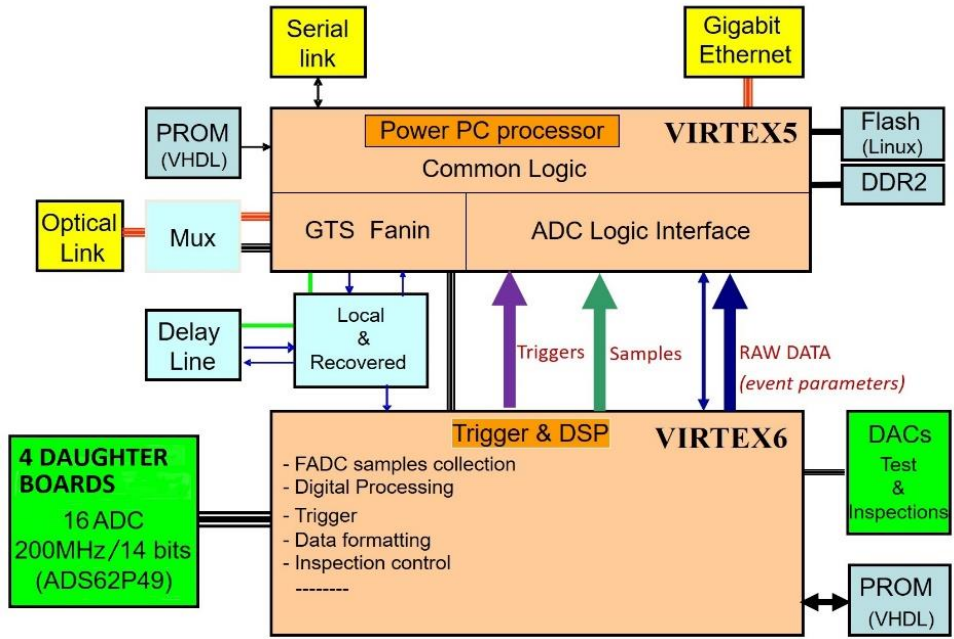

**Figure 1:** Block diagram of NUMEXO2 main board.

Each daughter card holds an HDMI (High-Definition Multimedia Interface) connector for Analog input signals.
Four chained Logic/analog inspections outputs and other complementary logic signals such as TRIG_IN, STOP, CLK, OR-TRIG_OUT are present on the front panel (Figure 2).

The Inputs/Outputs on the rear panel are :
- One Ethernet copper link for data transmission at 1Gbit Ethernet and for slow control,
- One Optical link for the GTS (Global Trigger and Synchronization) system based on SFP (Small Form factor Pluggable) transceiver,
- Four optical links through zarlink transceiver under PCIe 1.0 protocol at 400MB/s to increase data transfer.

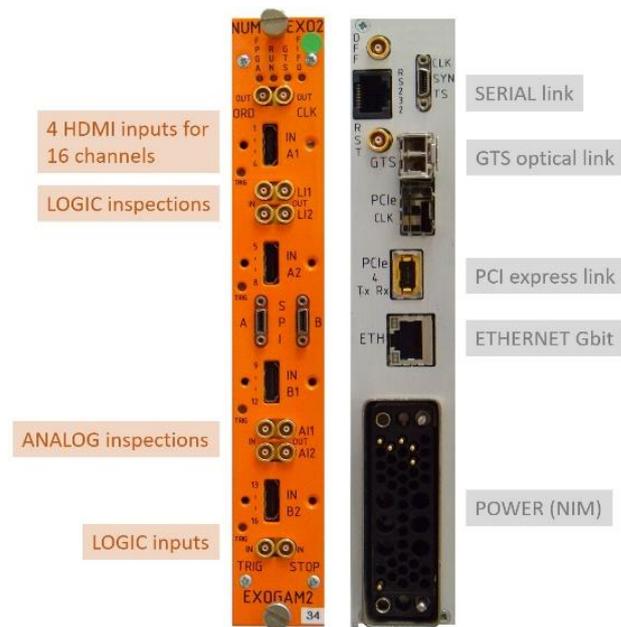

**Figure 2:** Front and rear panels description of NUMEXO2

## 3. Digital Signal Processing

### 3.1 General introduction

All of the Digital Signal Processing (DSP) algorithms of NUMEXO2 are implemented on the VIRTEX6. The originality of this work compared to similar systems is that all numerical operations follow the digital data flow from ADCs, without any storage step of samples, which would have consumed more resources. The memory requirements are therefore limited to the available resources on the FPGA for the digital filters (BRAM blocks) as well as the various stages of FIFOs concerning data sending. This allows better optimization of FPGA memory and allows the integration of the processing of the 16 independent channels.

#### 3.1.1 Data Flow

The samples are sent by the ADC daughter boards to the VIRTEX6 through 112 LVDS (Low Voltage Differential Signaling) differential pairs. It represents a constant flow of 44.8Gbit/s (see figure 3). Data are deserialized by ISERDES inputs inside the VIRTEX6. Fourteen ISERDES blocks are needed to get the 14 parallelized bits coming from the two channels (see figure 4).

The physics parameters are extracted from the input signals through different programmable filters. The data length is reduced by about a 1/1000 factor (dependent on the counting rate of the detector signal) keeping only the relevant parameters like energy and time. The calculated parameters are inserted into a normalized data format frame and the Gbits Ethernet link is used to transmit data outside the module (TCP/IP protocol).

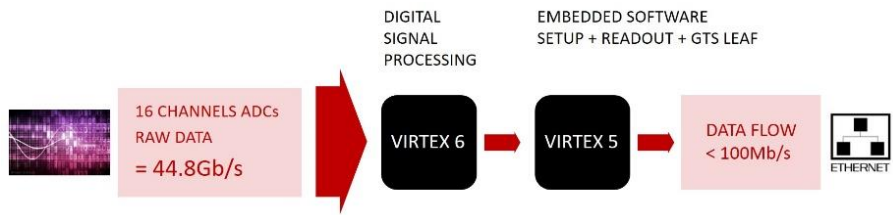

**Figure 3:** Data flow travelling from ADCs to the Ethernet Gigabit output link.

To ensure digital samples integrity, a control and calibration system for the 16 data lanes of 7-bit width is provided inside the VIRTEX6. The data are transmitted at 200MHz frequency by Double Data Rate method. The differential lanes on the PCB are length matched and parallelized to avoid time skew effects (length difference is around +/-5 mm corresponding to 70 ps).

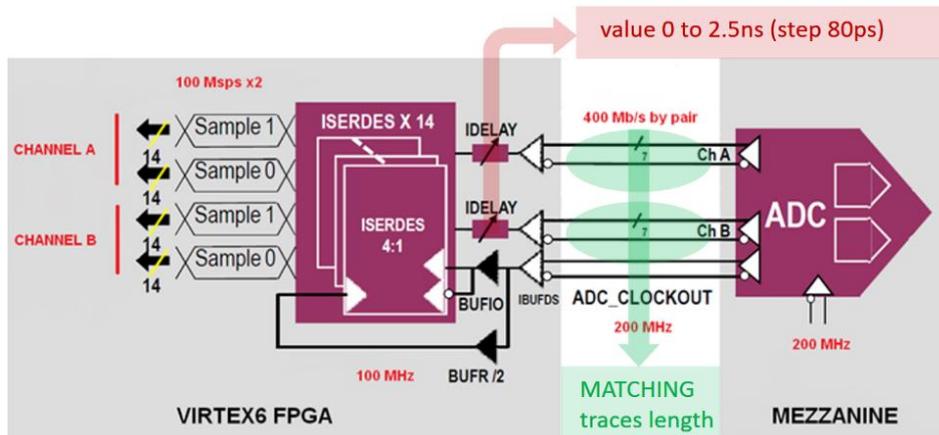

**Figure 4**: Block diagram of data flow between the ADC and the VIRTEX6 ISERDES inputs (for two channels A & B).

The 112 lines (7 lines per channel for 16 channels in total) with 400Mb/s data transfer (44.8 Gbits/s) inside the mother card PCB required a careful design and several signal integrity modeling and simulation using Cadence tools. In addition, we have developed a delay automation IP to optimize the delay lines compensation inside the VIRTEX6 on ISERDES Inputs. The optimal value search is based on the programmed bit pattern generated by the ADC and scanning all the possible delay values at 80ps step (IODELAY dedicated blocks) during a half memorization period (2.5ns). The result is a 32-bit word for each ADC channel in which the median value of the largest set of consecutive ones represents the optimal time to memorize data (see Figure 5).

Channel A

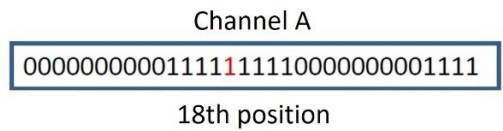

18th position

**Figure 5:** For instance, for channel A of the first ADC of the first daughter board, optimal position of delays found by the automation IP in 18th position from the right.

The memorization gate duration is in the (640-720ps) timing interval with a theoretical period of 2.29ns (including 210ps setup and hold times of memorization flip/flops) [5]. The optimal positions are stored inside an external EEPROM and automatically loaded on every power on.

### 3.1.2 Data frame

Each Channel is treated independently with its own Digital Signal Processing. A signal is validated for treatment if the analog signal coming from the detector is greater than a programmed threshold, which will generate a 10 ns trigger request signal. After full treatment, a data frame including several information such as the module and channel numbers and the physical measurement (the amplitude of the input signal and its arrival time), is generated in MultiFrame Metaformat. The main goal of the MultiFrame metaformat (MFM) is to be able to define binary formats for data acquisition and serialization that are self-contained, layered, adapted to network transfers and evolving.

### 3.1.3 Structure of Digital Signal Processing IPs

Digital Signal Processing IPs are developed under VHDL and Verilog languages. Verilog is easier for signed number manipulation and the VHDL is better for hierarchical structuring of the firmware. The firmware has up to five level of descriptions. The top level is composed of READOUT, DSP, INSPECTIONS and SETUP IPs (Figure 6). The massively parallel architecture of FPGA leads to the duplication of identical blocks up to 16 in our case. Synplify Pro from Synopsys [6] was used for logical synthesis since it offers a better schematic view than any other competing tools.

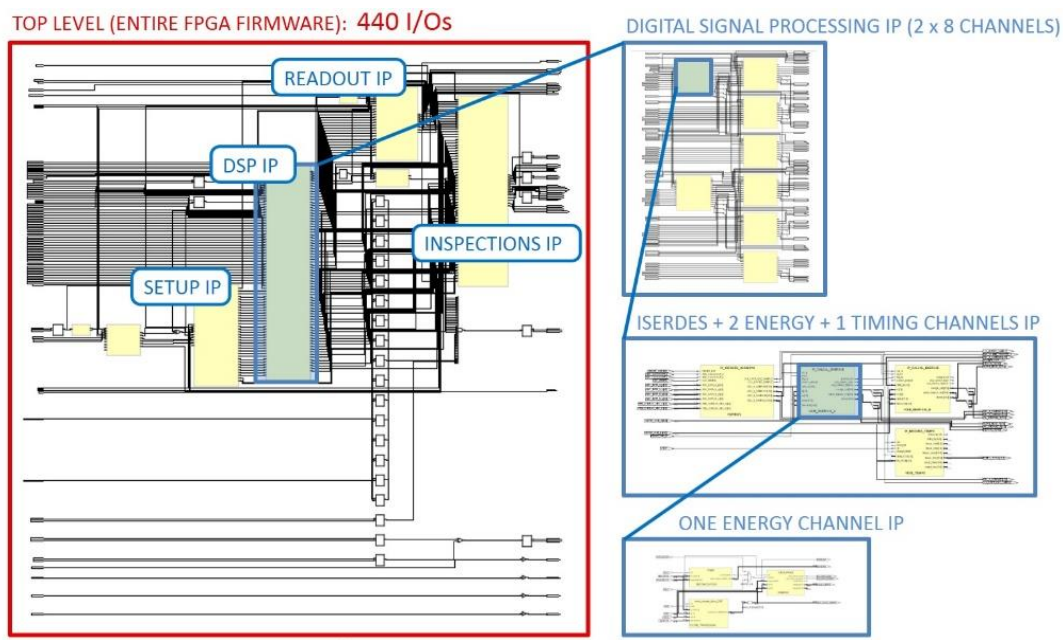

**Figure 6:** RTL (Register Transfer level) view showing the hierarchical structure of the VIRTEX6 firmware.

### 3.1.4 Data Transfer between FPGA

The reading and transfer mechanism is based on three types of FIFOs : two FIFOs per Channel and one global FIFO. The Channel dedicated FIFOs allow to store the calculated parameters after the signal treatment during one clock cycle. Then a full state machine encapsulates the data stored in the global FIFO in a specific MFM frame before transmission to the VIRTEX5. FIFO Flags and round robin methods are implemented to serve each of the 16 FIFOs of data frames independently of their counting rates.

The data transfer between the VIRTEX6 and the VIRTEX5 is based on an asynchronous protocol. Data (16 bits) are synchronized by DATA_STROBE signal and the acknowledgement is done by the ACK_LINK signal coming from VIRTEX5 (receiver). This protocol is based on the well-known VME standard. The Gigabit Ethernet output rate is limited to 10MB/s but it is sufficient to evacuate the frames at 200000 frames/s.

A BUSY signal is generated by the receiver (VIRTEX5) to control the data flow between the VIRTEX6 and the VIRTEX5. The transmitter (VIRTEX6) uses the BUSY signal to stop Trigger Request generation and the production of data frames.

### 3.2 Trigger and Timestamp

### 3.2.1 Trigger

The raw samples coming from the ADCs at 200 MSample/s (Msps) are processed at 100 Msps to match the GTS frequency, to produce an output signal according to the following equation:

$$S[n] = F[n] - \alpha \times F[n-1] \tag{1}$$

where F[n] is the filtered input signal (see 3.3.1) at sample n and S[n] the output signal. This equation corresponds to a differentiation filter where $\alpha$ is the decay constant of the charge preamplifier (around some microsecond). When the S[n] signal passes through a programmed threshold, a trigger (prompt) is produced and sent. The data are then transmitted to the next stages of NUMEXO2 (VIRTEX5).

To avoid the time walk on the output signal, a digital Constant Fraction Discriminator (dCFD) can also be used. The dCFD is calculated using the equation:

$$dCFD[n] = S[n-D] - F \times S[n] \tag{2}$$

The Delay D and the Fraction F are programmable by 10ns and 10% step, respectively. The trigger request signal is produced as a result of the zero crossing of the dCFD signal.

### 3.2.2 GTS tree and Time Stamp

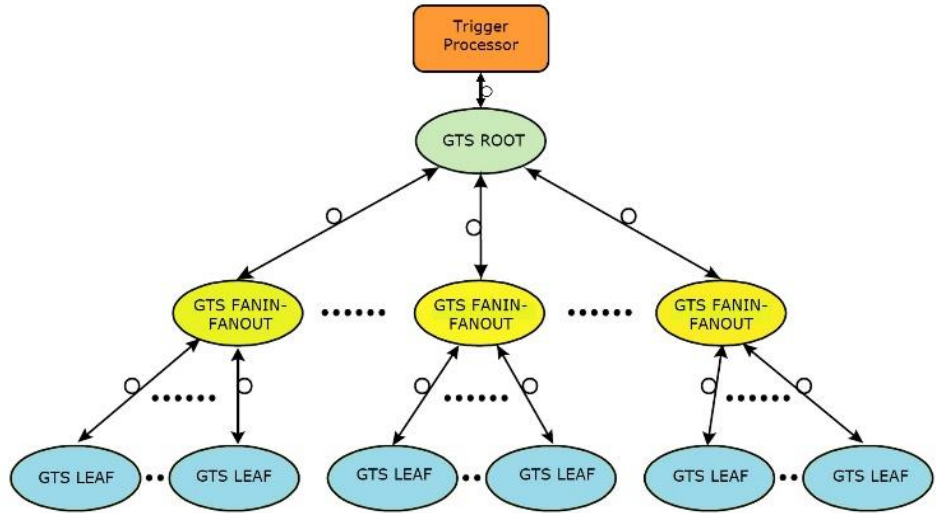

**Figure 7:** GTS (Global Trigger and time Stamp) tree.

The Global Trigger and time Stamp (GTS) system [3] can be viewed as tree composed of one ROOT and several leaves (the NUMEXO2 boards). The GTS tree allows the clock and Time Stamp distribution from the root to the leaves (Figure 7). At the leaf level, the 100 MHz clock is recovered from the root messages. The Time Stamp is a 48-bit wide counter incremented every 10ns. Each Trigger Request is Time Stamped. Thus, if a decision maker like a Trigger Processor is connected to the ROOT, each triggered leaf sends the channel identifier associated to its Time Stamp to the Trigger Processor that renders a decision and sends it back to the leaves. If the decision is an acceptance, data will be sent to the host computer otherwise the data is rejected. The GTS system is limited to generating and sending one Trigger Request per leaf. However, to enable triggering on all 16 channels of the NUMEXO2, the system stores information about which channel(s) triggered in a local memory. This approach ensures that when the trigger validation is finally sent back to the leaf, it is possible to determine which channel(s) is associated with it.

The GTS leaf IP, part of the VIRTEX5 firmware and managing the connection to the GTS tree via the optical fiber, assigns a TimeStamp to the channel at the arrival of the Trigger Request signal, supply by the DSP IP of the corresponding channel in the VIRTEX6. The Time Stamp is an absolute time label encoded over 48 bits with 10ns granularity transmitted by the ROOT board of the GTS tree. This labelling is fundamental in any experiment to correlate the channels belonging to the same physical event. The GTS leaf IP participates as well in the calibration operation used to align the tree according to the disparities in topology and length of the optical fibers linking the system's boards.

### 3.2.3 Test performed to check the alignment

As mentioned in **3.2.2**, it is essential in physics experiments to check that the data produced by detectors in time coincidence are correctly correlated. Generally, the solution consists in duplicating the detector signals and testing the correlation between the amplitude and the associated Time Stamp. To avoid these constraints another optional method has been developed and implemented in NUMEXO2. It relies on just 2 elements to check correlations at any time, whatever the count rate and the type of DSP. A very low frequency (< 10Hz) logic signal is sent to the TRIG IN inputs of the NUMEXO2 boards, which synchronously generates the automatic production of calibration data blocks competing with the real data from physics. These calibration data are identified by a "fictitious" energy value of 60,000 and the setting of the DNV status bit. These data allow a check of the Time Stamp homogeneity of all channels at + /- 10ns.

### 3.3 Filters

#### 3.3.1 Trigger filtering

In order to trigger at very low energy thresholds (~20keV over a range of 6MeV), it is essential to filter the response of the differentiation triggers, which considerably amplifies sudden variations, and therefore high frequencies. To this end, a cascade of three Infinite Impulse Response (IIR) low-pass filters with slightly offset cut-off frequencies is added (Figure 8). These first-order filters use very few resources (slices and DSP48), which is very advantageous given the number to be implemented (3x16) in the DSP48 IP of VIRTEX6. The recursive equation for a low pass filter is as follows :

$$F[n] = a \times E[n] - b \times F[n-1] \tag{3}$$

with E the input signal and F the ouput signal. The a and b values used for the three low pass filters are **a = 0.2, 0.3 and 0.4 and b = 0.8, 0.7 and 0.6.**

The maximum gain is obtained for 3.5MHz, i.e. for slow detectors as HPGe with a rise time (rt) of 100ns, a first-order equivalent frequency equal to 0.35/rt. An example on a real signal is given on figure 9.

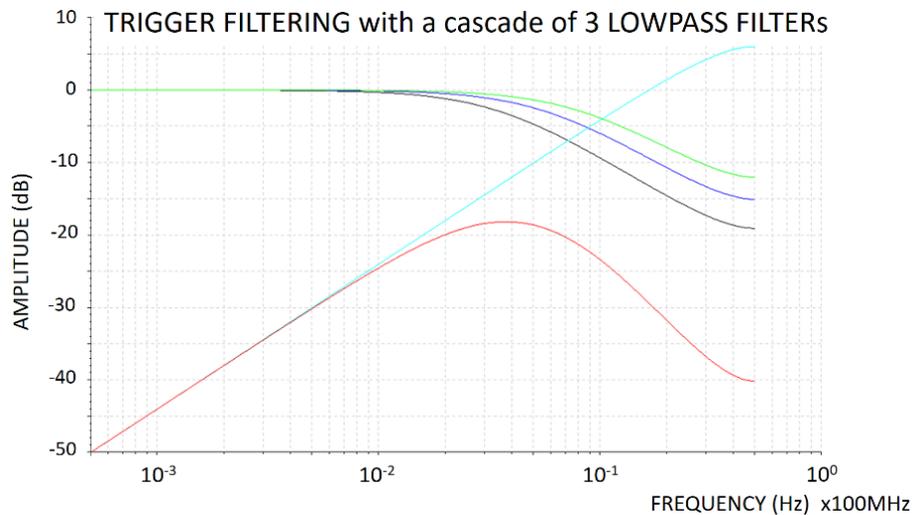

**Figure 8 :** The final frequency response of the filtered trigger function (red) after moderation by the cascade of LP filters (black, dark blue and green) and the trigger differentiation function (light blue, issued from equation 1).

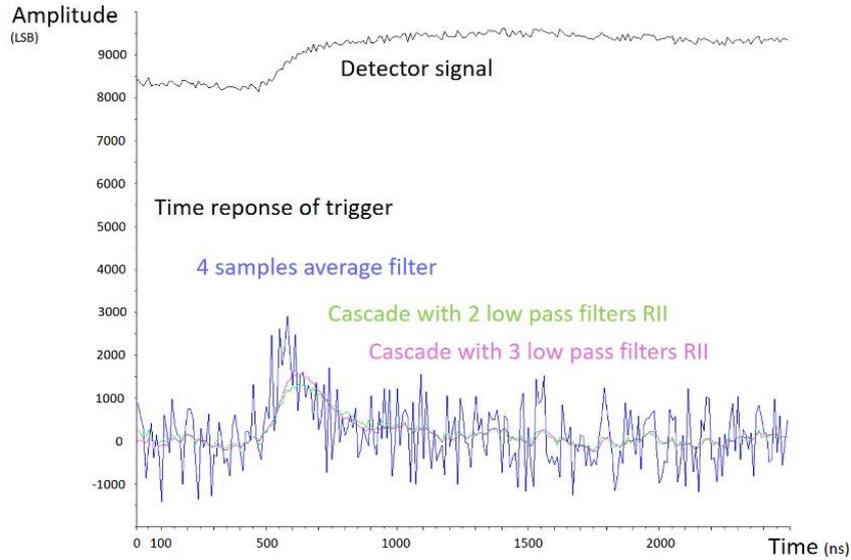

**Figure 9** : The time response of the trigger signal (pink) filtered by the cascade of the 3 LP filters and the trigger differentiation, a basic filtering for comparison (blue). The very low energy detector signal is shown in black.

### 3.3.2 Energy Filtering

The filter used to extract the energy value is of the trapezoidal type, described and formulated by Jordanov [7]. It is easier to implement it digitally than a Gaussian filter for equivalent measurement results. In addition, this filter is temporally bounded and is therefore more appropriate for high counting rates (10-100kHz) with a lot of pile-up signals. The raw samples of the input signals (E[n]) are fed directly into the filter, that calculates continuously at 100Msps according to the following equation:

$$T[n] = 2T[n-1] - T[n-2] + E[n-1] - \alpha E[n-2] - E[n-(k+1)] + \alpha E[n-(k+2)] - E[n-(k+m+1)] + \alpha E[n-(k+m+2)] + E[n-(2k+m+1)] - \alpha E[n-(2k+m+2)] \qquad (4)$$

where :
- E[n] is the input signal at sample n,
- T[n] is the ouput signal from the filter equation,
- k is the slope of the trapezoidal filter and m the duration of the flat top,
- and finally $\alpha = e(-Te/\tau)$ (always < 1) is a coefficient without unit with Te the sampling period equal to 10ns and $\tau$ the decay constant of the charge preamplifier signal.

For example with standard HPGe charge preamplifier : $\tau = 50\mu s$, the value of $\alpha$ is 0.9998.
The meanings of k, m and $\alpha$ are also reported latter on figure 11.

This recursive filter uses only 32-bit signed integers to calculate multiplications. Its implementation makes full use of the massively parallel structure of the chosen FPGA, exploiting the dedicated DSP multiplier blocks, i.e. the calculation of 400M multiplications/s per channel, 6.4G multiplications/s for the 16 channels. The graphical result after the logical synthesis (Figure 10) clearly shows all the filter elements (multipliers, final sum, circular buffers for delay).

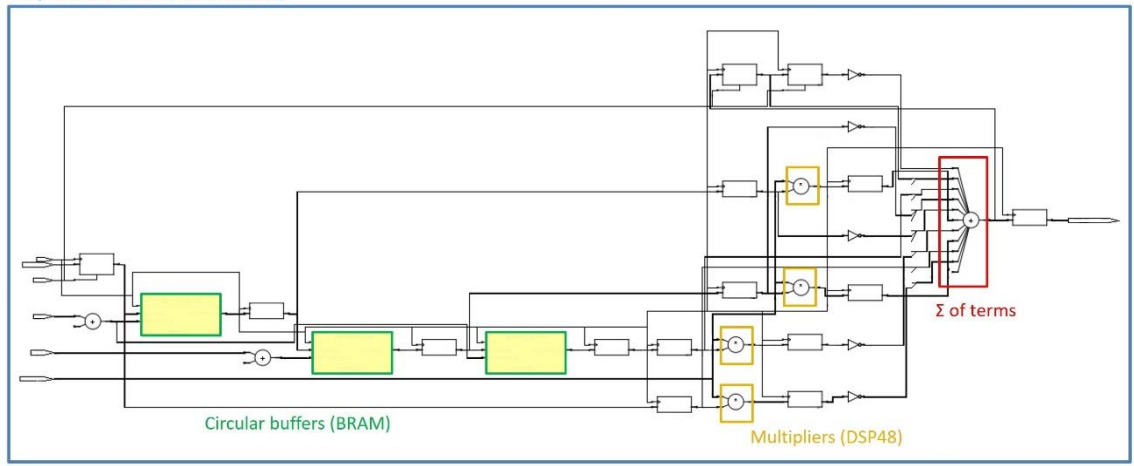

**Figure 10 :** RTL view of the implemented trapezoidal filter.

The k and m parameters, the duration of the rise time and the flat top of the trapezoidal respectively, are set so as to find the optimum signal to noise ratio and therefore optimum energy resolution, as a function of the detector noise contributions and expected count rate. To cancel low-frequency fluctuations in the baseline, the calculated energy value corresponds to the difference between the averages of the baseline and the amplitude of the flat top (Figure 11).

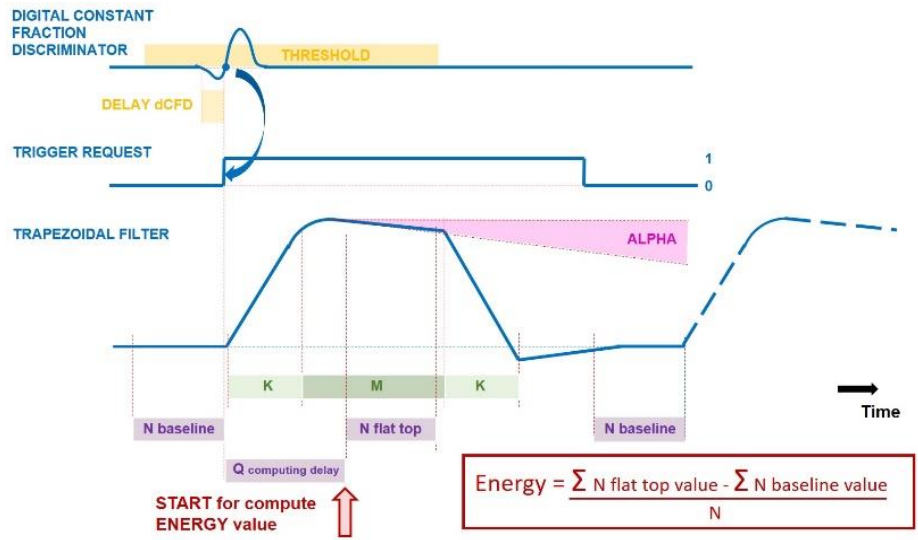

**Figure 11 :** schematic time diagram of the dCFD and trapezoidal filter showing all possible adjustable parameters.

The transfer function of the trapezoidal filter shows that the gain is proportional to the parameter k. Furthermore, the larger the k value, the better is the high frequency rejection (Figure 12).

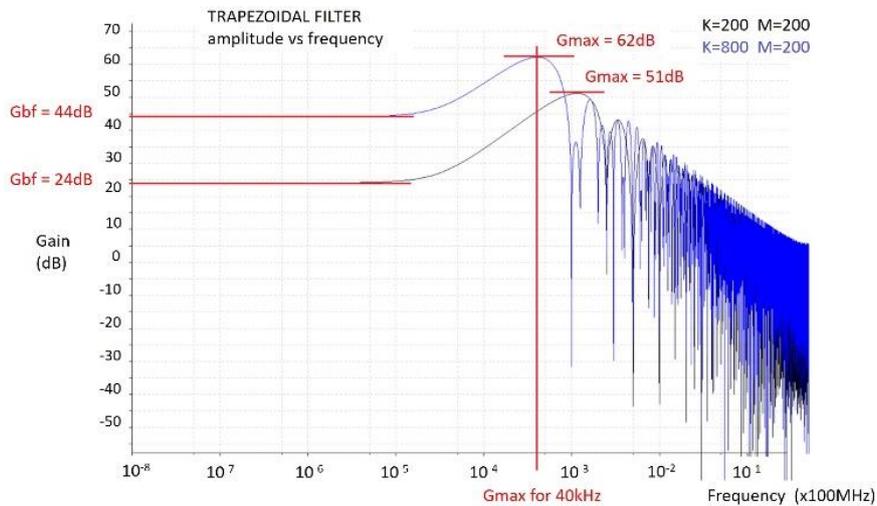

**Figure 12** : Frequency response of the trapezoidal filter for two values of the k parameter.

At high counting rate, pileup will affect the signal. Depending on the proximity between two trapezoidal filters a Data Valid (DV) or Data Not Valid (DNV) is produced by the DSP block. It is the conjunction of the trigger signal and the trapezoidal filter, which allows the energy to be calculated or not, although the two types of processing are carried out in parallel from the same samples of the detector signal. When two input signals are not too close in time, the energy can be calculated and the DV signal is produced. Conversely, when the energy cannot be determined the DNV signal is produced (Figures 13).

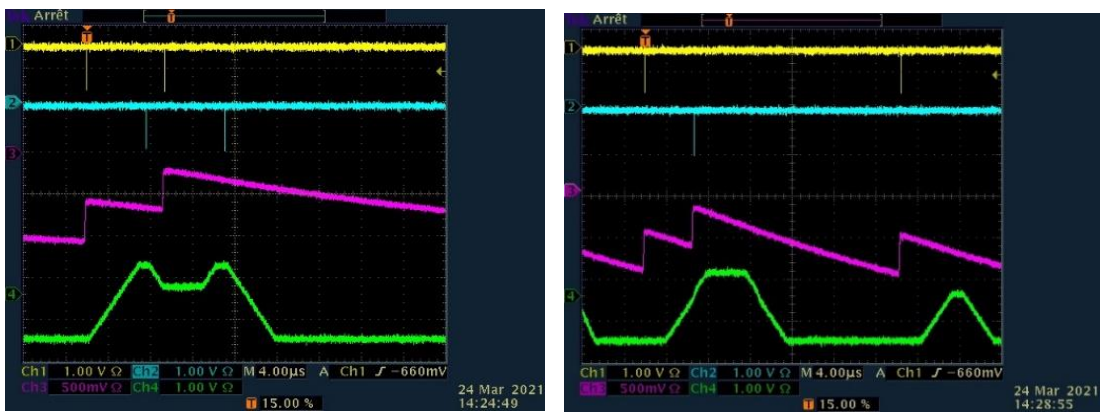

**Figures 13:** Example showing two calculated energies with Trigger Request signal (yellow), detector signal (violet), trapezoidal filters (green) and in blue (left) Data Valid and (right) Data Not Valid.

A dedicated logic validation gate coming from outside the NUMEXO2 can help to filter data. Only the Trigger Requests in coincidence with the validation gate will be processed.

### 3.4 Timing measurements

Like energy, time-of-flight (ToF Δt) measurement is an essential parameter for nuclear physics experiments. The intrinsic time resolution obtained is better than 1ns FWHM over a time range of 600ns, for a detector signal rise time of 100ns [8]. Given the performance to be achieved, the time-of-flight measurements (1 per channel), fully integrated in the FPGA, are based on the use of 2 very heterogeneous time measurements, firstly, by linear interpolation of the dCFD calculation (from equation 2) to calculate the Tstart, and secondly, by oversampling the common logic STOP signal arriving at the input STOP of the NUMEXO2 (Figure 14) to calculate the Tstop. This logic signal generally comes from a fast detector.

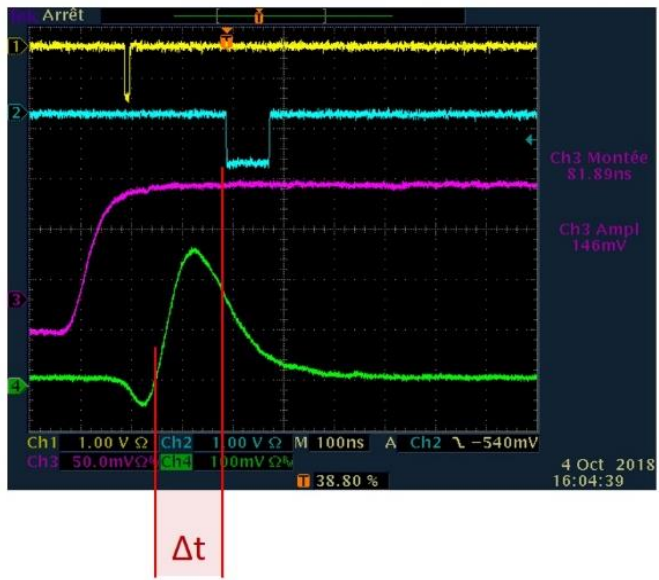

**Figure 14 :** Time diagram showing time of flight Δt measurement between TR (yellow coming from 0 crossing of green signal) and stop (blue), detector signal (pink) and dCFD (green).

### 3.4.1 Linear Interpolation of the dCFD signal

A time smaller than the sampling period (10ns) is obtained by calculating a linear interpolation in the 10ns time interval defined by the negative and positive values surrounding the zero crossing of the calculated dCFD signal (from equation 2). A Full State Machine (FSM) manages the search of the zero crossing time in this time interval by dichotomising it into 10 steps, lasting 50ns per step. The error is reduced by a factor of 2 at each calculation iteration, i.e. $10ns/2^{10} = 97ps$ of uncertainty at the end of the 10 steps. Nevertheless, the error introduced by the reduction in the number of effective quantisation steps (ENOB) (11.6-11.7) of the ADCs [9] is predominant for low amplitude values in the degradation of time resolution . This dichotomy solution has the advantage of a deterministic latency (500ns) and can be implemented for 16 channels, without the need of a divider block, which would be very resource-demanding (DSP blocks and slices). This calculation gives an acccurate Tstart in the 10ns period.

### 3.4.2 Oversampling of the STOP signal

The first edge of the STOP logic signal must be measured accurately. To this end, an oversampling method is implemented in the FPGA [10], using 4 clocks with a frequency of 400MHz derived from 100MHz, phase-shifted by 0°, 45°, 90° and 135° respectively and stored on all clock edges (DDR mode). The equivalent sampling frequency obtained for the STOP logic signal is therefore 3.2GHz (400MHz x 8), giving an uncertainty period of 312.5ps (Figure 15). Through several stages of FPGA flip-flops, the output result is a 32-bit pattern with a sequence of contiguous 1. The one-to-zero transition indicates the signal transition time, with a resolution of 312.5ps (10ns/32 steps).

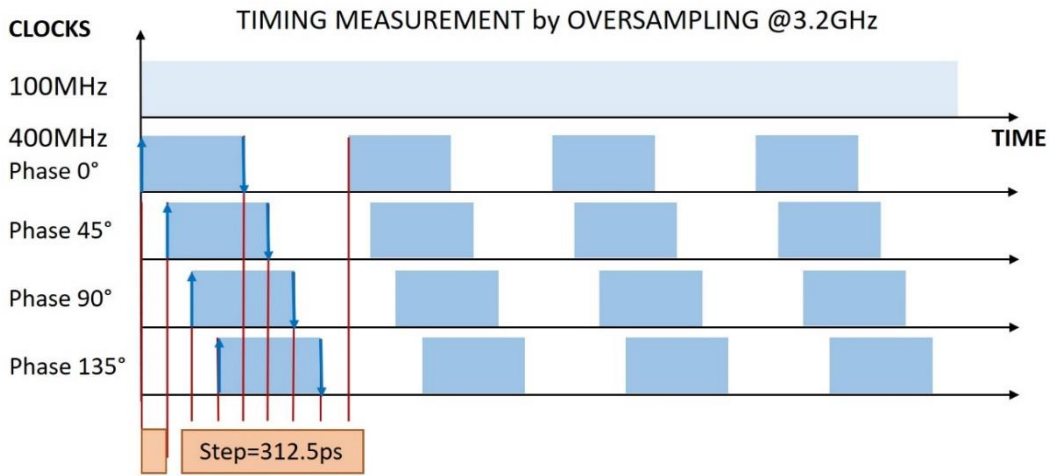

**Figure 15 :** Time diagram of oversampling method.

The oversampling stage must be strongly constrained in terms of placement to ensure an homogeneous and constant transit time whatever the future rerouting of the DSP. The placement of the first eight flip-flops (FDRE) receiving the common STOP signal is strategic and they are geographically constrained. Similarly, in order to optimize time measurement, the STOP signal is driven by a global clock buffer to take advantage of the low skew between the propagation times of this signal to the first flip-flop.

In addition, an internal calibration of the FPGA, that is necessary at the boot of the board, makes it possible to measure the DNL (Differential Non Linearity) in order to check for missing codes, represented by holes in the spectrum and to correct the inhomogeneities of the phase relationships produced by the MMCM (Mixed Mode Clock Manager) of the DSP/FPGA, managing the 4 clock phases. The random and equiprobable nature of the STOP signal in relation to the out-of-phase clocks means that it scans all 32 possible steps over 10ns (Figure 16). The spectrum shows a structure of phase relationships with a DNL of around 20%, i.e. a difference between phases of 60ps at most, to be compared with the intrinsic jitter equal to 90ps peak-to-peak at the output of the MMCM used.

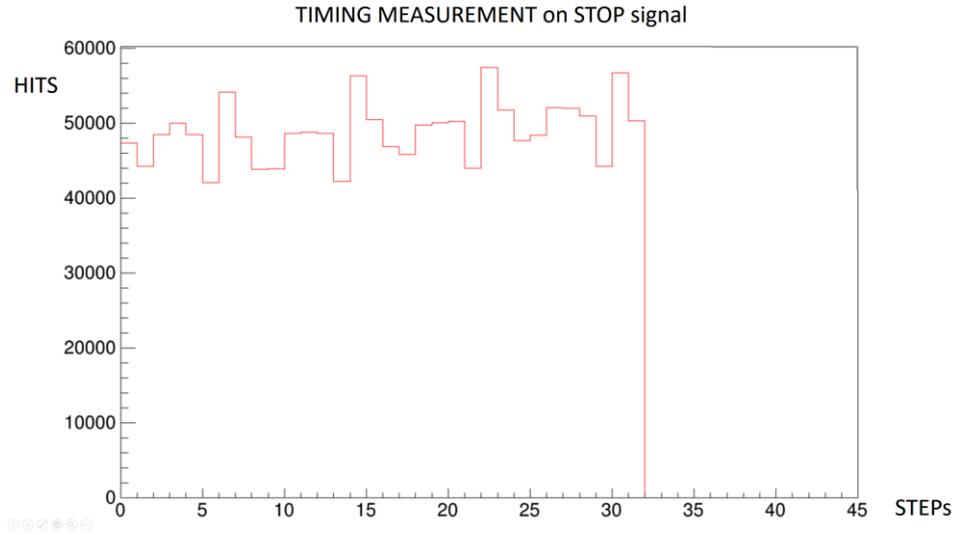

**Figure 16** : Spectrum of the Tstop measurement, as an image of the DNL, corresponding to 32 steps-inside 10ns period (312,5ps per step)

Finally, the time-of-flight measurement is issued from the following steps :
- Calculation of the linear interpolation between 2 successive amplitudes values at zero crossing of the dCFD (Tstart),
- The integer number of sampling periods of 10 ns (Tperiod) beetween Tstart and Tstop,
- The accurate value Tstop of the arrival time of the STOP signal in the 10 ns interval of the sampling period in number of 312.5ps steps.

The final formula for the ToF measurement $\Delta t$ is :

$$\mathbf{\Delta t} \text{ (in ns)} = \delta . \text{Ns} \qquad (6)$$

With $\delta$=10ns/32 i.e. 312.5ps,

And $\mathbf{Ns} = 32 \times (1024 - \mathbf{Tstart})/1024 + 32 \times \mathbf{Tperiod} + \mathbf{Tstop}$ $\qquad$ **(7)**

where Ns is a number of steps of 312.5 ps between Tstart and Tstop.

### 3.5 Charge Integration

Another use of DSP with NUMEXO2 is the energy measurement by summing the received samples contained in an integration window. This type of calculation uses the full NUMEXO2 sampling frequency of 200MHz. This type of processing is used for SiPM signals for example.

### 3.6 TAC digitization

For very fine time measurements, less than 500ps, NUMEXO2 can process a signal from an analogue Time to Analogue Converter (TAC) module. It is triggered by the signal to be measured and the FPGA performs a differential sum going from the baseline to the flat top zone of the TAC signal. The use of a recursive moving sum shifted by N samples (power of 2) is very efficient in terms of resources , without the need to store the samples and therefore without calculation latency. The used recursive moving sum formula, which makes a difference in the input signal to produce each point of the output signal, is as follows:

$$\mathbf{Sum[i]} = \mathbf{Sum[i-1]} + \mathbf{E[i]} - \mathbf{E[i-N]} \qquad (8)$$

with :
E[i] input samples at 100Msps,
Sum[i] the average values calculated every 10ns,
and N the shift.

The result is obtained by taking into account the significant bits of the Sum value.

### 3.7 Analog Demultiplexing Capabilities

For detectors with hundreds of channels, such as Multi Wire Proportional Chambers (MWPC), the signals are pre-processed by specific integrated circuits (ASICs). In the case of GASSIPLEX [11], they generate analogic pulse trains at the output, where the amplitude level is the image of the energy value. Logic sequencing signals (clock, track and hold logic signals) are required by the ASICs to store and output the processed signals. These can be generated by a CAEN VME V551 module. The NUMEXO2 card can manage 2048 channels, divided into 16 channels for a train of 128 amplitude values. It also uses the clock signal to sample and apply threshold (on the fly) to all the amplitude values. A data block specific to this function, of fixed size 280 bytes, is generated at the output for 1 to 64 data values above the threshold. However, for more data per event, one or more blocks are added, or all 32 data blocks for all 2048 channels. There is no dead time for processing or evacuating data blocks thanks to on-the-fly data processing and partly due to the 2 stages of FIFOs for frame readout.

## 4. Results

### 4.1 High Resolution γ-ray spectroscopy

The energy resolution was measured using an EXOGAM Ge detector with $^{60}$Co and $^{152}$Eu sources [1,12]. For the 1.3 MeV peak of cobalt, the measured energy resolution is 2.29 keV (Figure 17).

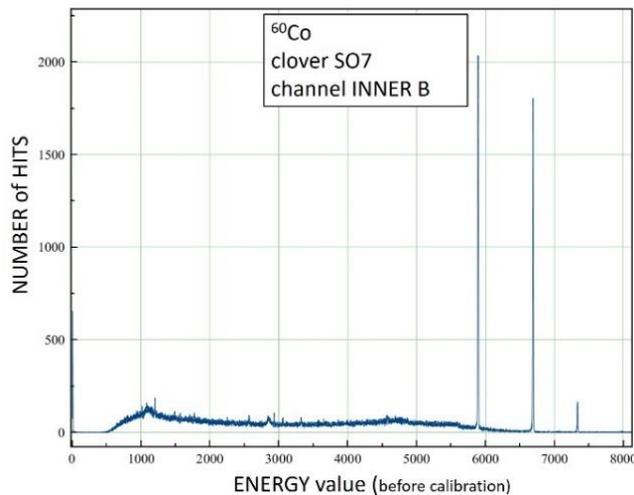

**Figure 17 :** $^{60}$Co source energy spectrum obtained from an INNER contact of an EXOGAM Ge detector.

With europium (Figures 18), the two peaks at 1085 and 1089 keV are clearly separated. The very low energy trigger threshold (< 30keV) allows the detection of X-rays at 39.9 keV and 45.7 keV respectively.

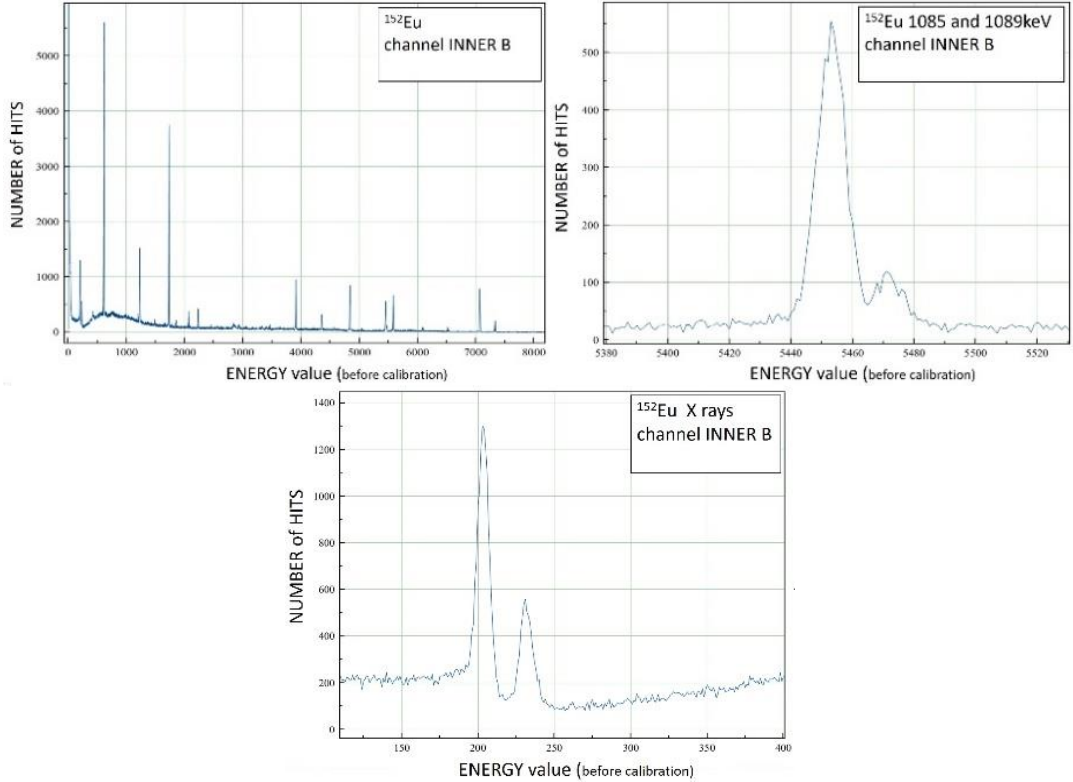

**Figures 18 :** $^{152}$Eu source energy spectrum obtained from an INNER contact of an EXOGAM clover : full scale (left) ; zoom on the 1085-1089 keV lines (middle) and zoom on the low energy part showing the X-rays.

### 4.2 Timing measurement

We measured the intrinsic time resolution of the NUMEXO2 board. It was carried out using a CAEN NDT6800D generator, using the detector-like signal from the analog output and the logic output for the STOP signal. Figure 19 shows a time resolution close to 500ps FWHM (1 LSB = 312ps) for a pulse generator rise time signal equals to 100ns.

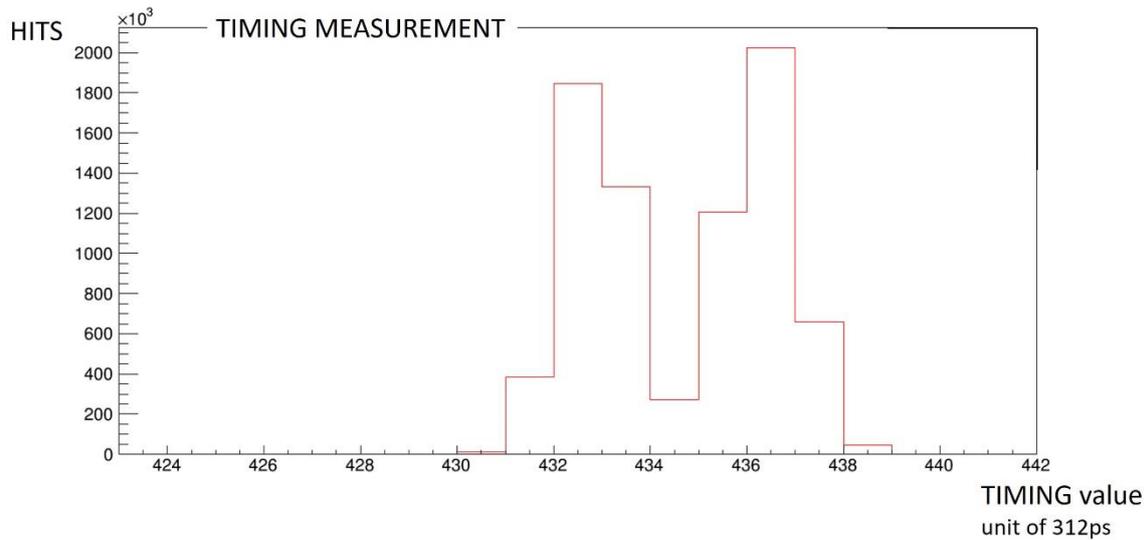

**Figure 19 :** Intrinsic time resolution obtained with a pulse generator. The second peak visible on the figure corresponds to the addition of a 1ns delay on the STOP signal via a 20cm coaxial cable.

This method is for instance used to perform time of flight measurements between a Silicon strip detector and a stop signal coming from fast gaseous detector.

## 4.3 High counting rate energy measurement

This measurement illustrates the ability of the NUMEXO2 DSP to perform energy measurements of random signals (Poisson distribution) with counting rates as high as 1Mevt/s per channel. The CAEN's NDT6800D pulse generator provides a combination of 3 distinct amplitudes, separated by 1mV for a median amplitude of 20mV, i.e. 1/1000 compared with the 2V of the full dynamic range of NUMEXO2 input stage. The trapezoid filter parameters k=100ns and m=200ns are constrained to absorb this counting rate while preserving a reduced dead time (see Figure 20). During this test, an external validation window during which NUMEXO2 board is active (NUMEXO2 input TRIG_IN, common to all channels) selects the number of frames to be output by NUMEXO2 to reduce the output data flow to a maximum of 6MB/s (approximately 200K frames/s). It is usually used in practise during experiment using the coincidence of two detector channels for instance.

The processing dead time with NUMEXO2 is linked solely to the sum of the durations of the k+m parameters of the trapezoidal filter, i.e. 300ns, although the discriminator also has its own minimum dead time equal to 50ns. The 48-bit counting scales integrated into NUMEXO2 show that 85.5% of triggers lead to a valid energy calculation for a counting rate of 1Mevt/s, and this number rises to 98.6% for 100kevt/s. The amplitude resolution is 0.043% over the dynamic range of 1V. These performances are particularly important for detector working at high couting rate like large gamma-ray arrays (EXOGAM) or in-beam ionization chambers.

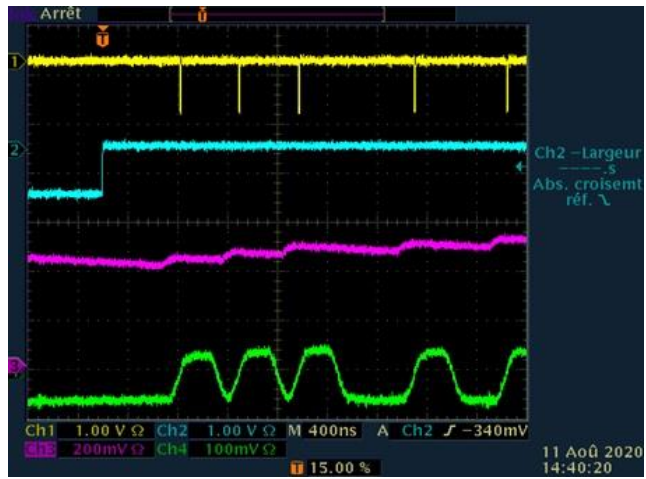

**Figure 20 :** High counting rate capability of NUMEXO2. The shown signals are: TR (yellow), external validation window (blue)**,** pulser (pink), trapezoidal filter (green).

Figure 21 shows the energy spectrum for almost 400 million events, showing three amplitudes separated by 1mV. The three corresponding peaks are clearly visible even  at 1 Mevt/s  counting rate.

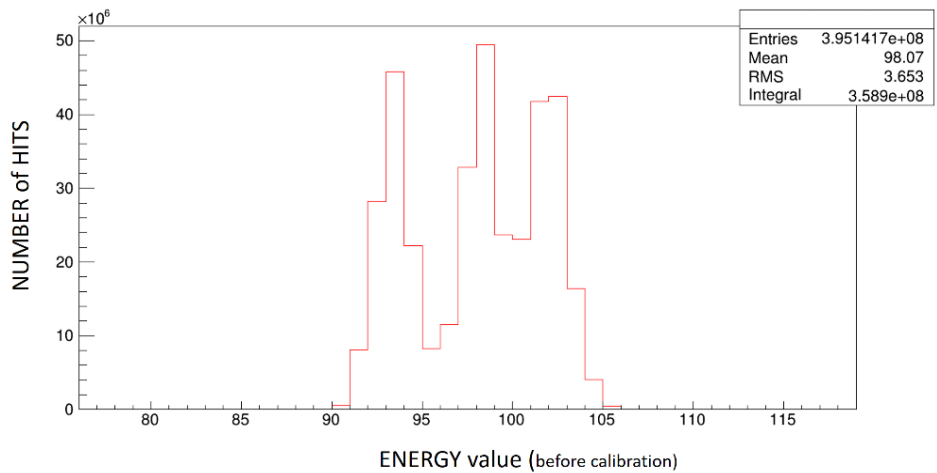

**Figure 21:** Energy spectrum obtained at 1 Mevt/s for three signals with an amplitude difference of 1 mV.

# 5. Conclusion

In this paper, we have described the NUMEXO2 boards in use at GANIL. While initially developed to process signals from HPGe detectors, NUMEXO2 is now operated with many different kind of detectors : HPGe, Si DSSD, Ionisation chamber, plastic scintillators, Multi Wire Proportional Counter, drift chamber, etc. The use of FPGAs and dedicated firmware configurations allows this extreme flexibility whilst preserving optimal performances for each detection system. With a few examples, we have shown that in each case the required energy and time resolutions were achieved. Even at very high counting rates the dead time is limited to the trapezoidal filter parameters k+m convoluted to the dCFD intrinsic dead time which allows rates up to 1 MHz with very small losses. In addition, the general use of NUMEXO2 boards together with the GTS clock distribution system greatly facilitates the coupling of several detectors which is today a standard feature in experimental nuclear physics. In order to improve coupling capabilities for instance with other systems, a new system SMART (Sfp connectivity and MicroTCA for Advanced Remote Trigger) is currently under design at GANIL to replace the GTS. The current limitations of NUMEXO2 are twofold : the obsolescence of certain components (memories in particular) and the limited ADC sampling frequency (200 MHz). This latter is not adapted to the fastest detectors (PMTs, time signal from a gas detector, current signals, SiPM…) for which a GHz frequency would be required.

**Acknowledgements:**

This work was partially fund by MCIN/AEI/10.13039/501100011033 (Spain) with grants PID2020-118265GB-C4, PID2023-150056NB-C4, PRTR-C17.I01, by Generalitat Valenciana (Spain) with grant CIPROM/2022/54, ASFAE/2022/031 and by the EU NextGenerationEU and FEDER funds.